\documentclass[a4paper,11pt]{article}
\pdfoutput=1 

\usepackage[T1]{fontenc} 
\usepackage{amsmath,amssymb,mathtools}
\usepackage{color}
\usepackage{cite}
\usepackage[colorlinks=true,linkcolor=blue, citecolor=red, urlcolor=blue, bookmarks]{hyperref}
\usepackage{graphics}
\definecolor{comments}{RGB}{220,20,60}
\usepackage[text={17.1cm,24.6cm},centering]{geometry} 
\usepackage{braket}
\usepackage{comment}
\usepackage{tikz}

\usepackage{authblk}
\usepackage{cleveref}
\usepackage{multirow,makecell}
\usepackage{textcomp}
\usepackage{wasysym}
\numberwithin{equation}{section}

\newcommand{\beq}{\begin{equation}}
\newcommand{\eeq}{\end{equation}}
\newcommand{\bea}{\begin{eqnarray}}
\newcommand{\eea}{\end{eqnarray}}

\newcommand{\nc}{\newcommand}
\nc{\ir}{\mathrm{i}}
\nc{\dd}{\mathrm{d}} 
\nc{\eE}{\mathsf{e}}
\nc{\Tr}{\text{Tr}}
\nc{\id}{\mathbb{I}}
\nc{\tdet}{\tilde{\det}}
\nc{\J}{\mathcal{J}}
\nc{\B}{\mathcal{B}}
\nc{\Gammax}{\Gamma_{\mathrm{x}}}

\definecolor{text}{RGB}{0,0,0}
\definecolor{details}{RGB}{124,133,255}

\begin{document} 

\title{\bf Analytical results for the entanglement dynamics of disjoint blocks in the XY spin chain}

\author[1]{Gilles Parez  \footnote{ {\normalsize 
\texttt{gilles.parez@umontreal.ca}}}}
\affil[1]{\it Centre de Recherches Math\'ematiques (CRM),
Universit\'e de Montr\'eal,
P.O. Box 6128, Centre-ville Station,
Montr\'eal (Qu\'ebec), H3C 3J7,
Canada
}

\author[2]{Riccarda Bonsignori}
\affil[2]{\it Ru\dj er Bo\v{s}kovi\'c Institute, Bijeni\v{c}ka cesta 54, 10000 Zagreb, Croatia}
\date{October 10, 2022}
\maketitle

\begin{abstract}

The study of the dynamics of entanglement measures after a quench has become a very active area of research in the last two decades, motivated by the development of experimental techniques. However, exact results in this context are available in only very few cases. In this work, we present the proof of the quasiparticle picture for the dynamics of entanglement entropies for two disjoint blocks in the XY chain after a quantum quench. As a byproduct, we also prove the quasiparticle conjecture for the mutual information in that model. Our calculations generalize those presented in [\href{https://journals.aps.org/pra/abstract/10.1103/PhysRevA.78.010306}{M. Fagotti, P. Calabrese, Phys.  Rev. A {\bf 78}, 010306 (2008)}] to the case where the correlation matrix is a block-Toeplitz matrix, and rely on the multidimensional stationary phase approximation in the scaling limit. We also test the quasiparticle predictions against exact numerical calculations, and find excellent agreement. In the case of three blocks, we show that the tripartite information vanishes when at least two blocks are adjacent.

\end{abstract}

\baselineskip 18pt
\thispagestyle{empty}
\newpage

\tableofcontents

\section{Introduction}

Understanding the nonequilibrium dynamics of quantum many-body systems is a question that has been at the heart of quantum mechanics since its early days \cite{vN-29}. For the last twenty years, this field of research has experienced a renewed interest, due to groundbreaking cold-atom and ion-trap experiments that managed to simulate the unitary evolution of closed systems on long time scales \cite{gmhb-02,kww-06,cbpesfgbkk-06,tcfmseb-12,gklkrs-12}, as well as an intense theoretical activity \cite{CEM}. A key aspect is that quantum entanglement and its dynamics out of equilibrium play a large role in our understanding of fundamental problems, such as the equilibration and thermalisation of isolated many-body systems \cite{PolkonikovRMP11, GE15, DKPR15, CEM,EF16,cdeo-08} and the emergence of thermodynamics \cite{C:18, DLS:13, SPR:11,ckc-14}. Moreover, several experiments managed to measure entanglement-related quantities \cite{kaufman-2016,exp-lukin,brydges-2018,ekh-20,neven2021symmetry,vitale2022symmetry,rath2022entanglement}.

The simplest protocol to drive a quantum system out of equilibrium is known as a \textit{quantum quench}~\cite{cc-06,cc-07}. Consider a system described by a Hamiltonian $H(\lambda)$, where $\lambda$ is a set of external parameters. In the quench protocol, the system is prepared in the groundstate of some initial Hamiltonian $H(\lambda_0)$, and at time $t=0$ the parameters are quenched from $\lambda_0$ to $\lambda_1$ such that $[H(\lambda_0),H(\lambda_1) ]\neq 0$. For $t>0$, the system evolves unitarily under the action of the post-quenched Hamiltonian $H(\lambda_1)$. Because the two Hamiltonians do not commute, the subsequent dynamics is non-trivial. In particular, the investigation of entanglement dynamics after a quantum quench received considerable attention over the last two decades \cite{fc-08,ac-17,ac-18,c-20}. 

For a quantum many-body system in a pure state $|\psi\rangle$, one is typically interested in the entanglement between a subsystem $A$ and its complement, traditionally denoted $B$. The so-called \textit{R\'enyi entropies} are defined from the reduced density matrix $\rho_A=\Tr_B (|\psi\rangle \langle \psi |)$ as
\begin{equation}
\label{eq:Sn}
    S_n^A = \frac{1}{1-n} \log \Tr \rho_A^n.
\end{equation}

The limit $n\to 1$ of the R\'enyi entropies yields the celebrated \textit{entanglement entropy} \cite{bennett1996concentrating}. It is the von Neumann entropy of the reduced density matrix of subsystem $A$, namely
\begin{equation}
    S_1^A \equiv \lim_{n\to 1}S_n^A = -\Tr (\rho_A \log \rho_A).
\end{equation}

R\'enyi entropies quantify the entanglement between $A$ and $B$, irrespective of the geometry of $A$. In many cases considered in the literature, $A$ is a connected spatial region, such as a segment in a one-dimensional chain. However, the case where $A$ consists of disjoints blocks has also generated interest \cite{calabrese2010entanglement,fagotti2010entanglement,cct-09,atc-10,ip-10,coser2014entanglement}. Let us consider the situation where $A=A_1 \cup A_2$ consists of two blocks $A_1$ and $A_2$. From the entanglement entropies of the two blocks $S_n^{A_1\cup A_2}$, one defines the mutual information as
\beq
\label{eq:MI}
I_1^{A_1:A_2}=S_1^{A_1}+S_1^{A_2}-S_1^{A_1\cup A_2},
\eeq
as well as the R\'enyi mutual information
\beq
\label{eq:RMI}
I_n^{A_1:A_2}=S_n^{A_1}+S_n^{A_2}-S_n^{A_1\cup A_2}.
\eeq
Here, $S_n^{X}, \ X=A_1, A_2,A_1 \cup A_2$, is the R\'enyi entropy from Eq. \eqref{eq:Sn} where the reduced density matrix is obtained by tracing out the degrees of freedom of the complement of $X$ from the total pure-state density matrix $\rho=|\psi\rangle \langle \psi|$. The mutual information is not \textit{per se} an entanglement measure between $A_1$ and $A_2$ because it also contains classical correlations \cite{wvhc-08,cct-13}, and a proper measure of entanglement in that context is instead the entanglement negativity \cite{vw-02}. However, both the mutual information and the negativity share important properties, both in \cite{liu2022entanglement} and out of equilibrium \cite{ac-19,ac2-19,bertini2022entanglement}, and we relegate the study of the negativity to forthcoming investigations.

For a tripartite system $A=A_1\cup A_2 \cup A_3$, a relevant quantity that characterizes multipartite entanglement is the tripartite information $I_n^{A_1:A_2:A_3}$. It is defined as \cite{cerf1998information} 
\begin{equation}
\label{eq:I3}
    I_n^{A_1:A_2:A_3}= I_n^{A_1:A_2}+I_n^{A_1:A_3}-I_n^{A_1:(A_2\cup A_3)}
\end{equation}
and measures the extensiveness of the mutual information. In particular, a negative tripartite information indicates multipartite entanglement, and it is related to quantum chaos and scrambling \cite{hosur2016chaos,schnaack2019tripartite,sunderhauf2019quantum}. In the context of two-dimensional systems where $A_1,A_2,A_3$ are adjacent regions, the tripartite information coincides with the celebrated topological entanglement entropy \cite{kitaev2006topological}. Very recently, the tripartite information was investigated in the context monitored spin chains \cite{carollo2022entangled} and quantum quenches \cite{maric2022universality}. While it vanishes at all times for many quench protocols, the authors of Ref. \cite{maric2022universality} show that in some cases its dynamics yields universal information about the system.

In integrable models, the quasiparticle picture \cite{CC04,cc-05, ac-17,ac-18,c-20} describes the growth of entanglement in time after a global quench, in terms of ballistic propagation of pairs of quasiparticles of opposite momentum, that spread entanglement and correlations through the system. In the case of disjoint subsystems, the quasiparticle picture can be adapted to describe the dynamics of the mutual information \cite{ac-19} and the entanglement negativity \cite{mac-21}. The quasiparticle picture also  describes the growth of symmetry-resolved entanglement \cite{PBC21,pbc-21bis,pbc-22}, and it has recently been generalized to the case of dissipative free fermionic and bosonic systems \cite{ac-21, ca-22, ac-22,ac2-22,caceffo2022entanglement}.

While the quasiparticle picture provides impressive quantitative results that have been checked extensively through numerical investigations, \textit{ab initio} and analytical results for the entanglement dynamics after a quench remain scarce in the literature. In a seminal paper, Fagotti and Calabrese computed the complete time dependence of the entanglement entropy of a single interval after a quench in the XY chain \cite{fc-08}. Their derivation used the Toeplitz-matrix representation of the correlation matrix and multidimensional phase methods. However, a similar derivation for the case of disjoint blocks is still lacking. We also mention recent exact results for the negativity dynamics in dissipative models \cite{ac2-22,caceffo2022entanglement} and quantum circuits \cite{bertini2022entanglement}

In this paper, we provide the analytical derivation of the quasiparticle picture for the entanglement entropies of two disjoint blocks in the XY chain in presence of transverse field, generalizing the results of Ref. \cite{fc-08}. As a byproduct, this allows us to prove the quasiparticle picture result for the mutual information. Moreover, we show that the tripartite information vanishes at all times for the quench protocol we consider, and argue more generally that the quasiparticle picture implies a vanishing tripartite information. Finally, we mention that these calculations for the entropy dynamics of disjoint blocks have already been advertised and used in the context of symmetry-resolved entanglement measures \cite{pbc-21bis,pbc-22}. 

This paper is organized as follows. In Sec. \ref{sec:model} we introduce the XY model and express the related entanglement measures in terms of the two-point correlation matrix after a quench. We discuss the quasiparticle conjecture for the entanglement dynamics in Sec. \ref{sec:QPP}, and give the analytical proof for the case of two disjoint blocks in Sec. \ref{sec:proof}. We conclude in Sec. \ref{sec:conclusion} with a summary of the results and byproducts of our proof, and discuss further research directions.

\section{The XY spin chain and entanglement dynamics}
\label{sec:model}

In this section we review the XY spin chain and its diagonalization. We also specify the quench protocol under consideration and recall the definition of the R\'enyi entropies and mutual information in terms of the two-point correlation matrix. Finally, we review the quasiparticle picture for the entanglement dynamics, both in the case where the subsystem consists of one or multiple blocks of contiguous sites. 

\subsection{The quenched XY spin chain}
\label{subs:XY}

We consider the spin-$1/2$ XY spin chain in a transverse magnetic field with periodic boundary conditions. The Hamiltonian is 
\beq
\label{eq:HamiltonianXY}
H(\gamma, h)=-\sum_{j=1}^L \left( \frac{1+\gamma}{4}\sigma_{j}^x\sigma_{j+1}^x+\frac{1-\gamma}{4}\sigma_{j}^y\sigma_{j+1}^y +\frac{h}{2} \sigma_j^z \right),
\eeq
where $\gamma$ is the anisotropy parameter, $h$ is the external magnetic field, $L$ is the size of the system and $\sigma_j^{\alpha}$ are the Pauli matrices at site $j$. The model can be solved via the Jordan-Wigner transformation, which maps the system into a free Fermi gas in the presence of an external potential \cite{lsm-61}. 

We consider the following quench protocol. At time $t=0$, the system is prepared in the groundstate $|\psi_0\rangle$ of the Hamiltonian $H(\gamma_0, h_0)$ with some initial anisotropy and magnetic field $\gamma_0$ and $h_0$. The parameters are then quenched to the values $\gamma$ and $h$, and for $t>0$ the time-evolved state is 
\begin{equation}
    |\psi(t)\rangle  = \eE^{-\ir t H(\gamma,h)}|\psi_0\rangle. 
\end{equation}

In order to study the time evolution of the entanglement measures, it is useful to introduce the Majorana operators
\beq
\label{eq:MajoranaJW}
a_{2j-1}=\left(\prod_{k=1}^{j-1}\sigma_k^z \right)\sigma_j^x, \hspace{1.5 cm} a_{2j}=\left( \prod_{k=1}^{j-1}\sigma_k^z\right)\sigma_j^y,
\eeq
that satisfy
\beq
\label{eq:Majorana}
a_{2j-1}=c_j+c_j^{\dagger}, \quad a_{2j}=\ir (c_j-c_j^{\dagger}), \quad \{ a_m,a_n\}=2 \delta_{m,n},
\eeq
where $c_j,c_j^{\dagger}$ are the canonical spinless fermionic operators. For a subsystem $A$ consisting of~$\ell$ contiguous spins, the time-dependent correlation matrix $\Gamma_A(t)$ is a $2 \ell \times 2 \ell$ matrix built from $2 \times 2$ blocks \cite{cc-05} as
\beq
\label{eq:corrGamma}
\langle \psi(t) |\begin{pmatrix}
a_{2m-1} \\ a_{2m}
\end{pmatrix} \cdot \begin{pmatrix} a_{2n-1} & a_{2n} \end{pmatrix} | \psi(t)\rangle =\delta_{m,n}+\ir[\Gamma_A(t)]_{m,n}, \hspace{1 cm} 1 \leqslant m,n\leqslant \ell.
\eeq
More explicitly, the correlation matrix is a block-Toeplitz matrix
\beq
\label{eq:Toeplitz}
\Gamma_A(t)=\begin{pmatrix}
\Pi_0& \Pi_1 & \cdots & \Pi_{\ell-1}\\
\Pi_{-1} & \Pi_0 &  &  \vdots\\
\vdots & &\ddots &\vdots \\
\Pi_{1-\ell} &\cdots &\cdots &\Pi_0
\end{pmatrix}, \hspace{2 cm} \Pi_j= \begin{pmatrix} -f_j & g_j\\
-g_{-j} & f_j\end{pmatrix}.
\eeq
In the large-$L$ limit, $f_j$ and $g_j$ read \cite{fc-08}
\beq
\label{eq:fg}
\begin{split}
  f_j&=\ir \int_{-\pi}^{\pi} \frac{\dd k}{2 \pi} \eE^{-\ir k j} \sin \Delta_k \sin (2 \epsilon_k t),  \\
  g_j&= \int_{-\pi}^{\pi} \frac{\dd k }{2 \pi} \eE^{-\ir k j}\eE^{- \ir \theta_k} \left(\cos \Delta_k+\ir \sin \Delta_k \cos (2 \epsilon_k t) \right),
\end{split}
\eeq
with
\beq
\label{eq:quenchparameters}
\begin{split}
    \epsilon_k^2&=(h-\cos k)^2+\gamma^2 \sin^2k, \qquad \epsilon_{0,k}^2=(h_0-\cos k)^2+\gamma_0^2 \sin^2k,\\
    \eE^{-\ir \theta_k}&=\frac{\cos k-h-\ir \gamma \sin k}{\epsilon_k},\\
    \cos \Delta_k &=\frac{hh_0-\cos k(h+h_0)+\cos^2k+\gamma \gamma_0\sin^2 k}{\epsilon_k\epsilon_{0,k}},\\
    \sin \Delta_k& = -\sin k \frac{\gamma h_0-\gamma_0 h-\cos k(\gamma-\gamma_0)}{\epsilon_k \epsilon_{0,k}}.
\end{split}
\eeq

\subsection{Entanglement entropies and mutual information for disjoint intervals}
\label{subs:MI}

\begin{figure}[t]
\centering
\includegraphics[width=0.8\textwidth]{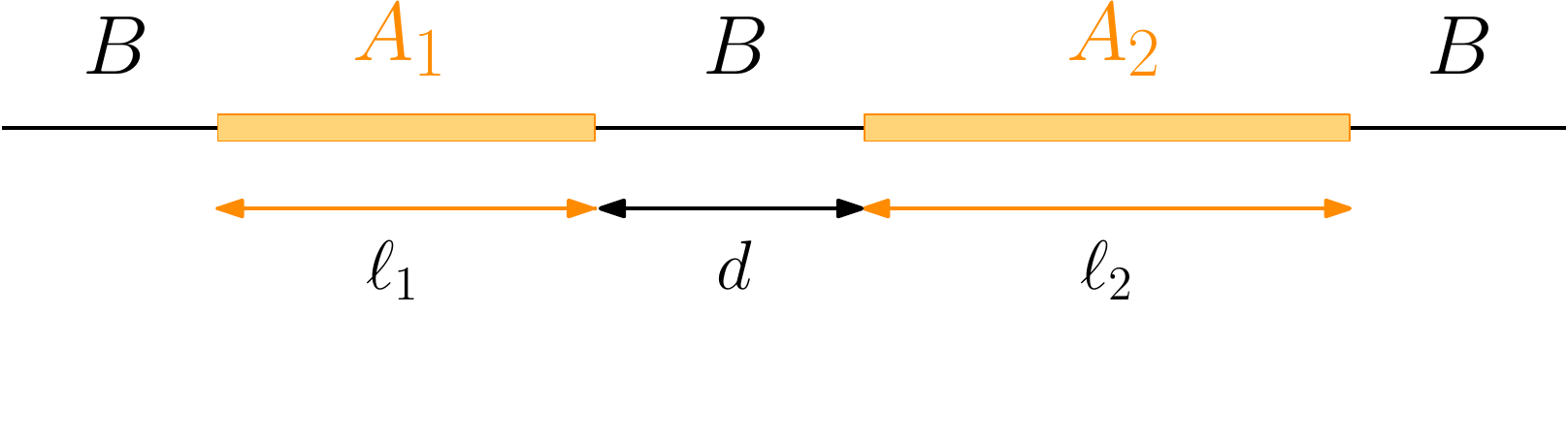}
\caption{The subsystem $A$ considered consists into two disjoint subsystems $A_1$ and $A_2$ of lengths $\ell_1$ and $\ell_2$, separated by a distance $d$.}
\label{fig:Int}
\end{figure}

Owing to the quadratic nature of the XY Hamiltonian in terms of fermion operators, the time-dependent reduced density matrix $\rho_A(t)$ of any subsystem $A$ is a Gaussian operator. This implies that the spectrum of $\rho_A(t)$ is related to the correlation matrix $\Gamma_A(t)$ associated to the same subsystem \cite{pc-01,p-03}, namely
\beq
\label{eq:Trrhocorr}
\Tr \rho_A^n=\left(\det \left[\left( \frac{\mathbb{I}+\ir \Gamma_A}{2}\right)^n+\left( \frac{\mathbb{I}-\ir \Gamma_A}{2}\right)^n \right]\right)^{1/2},
\eeq
where we removed the explicit time dependence for clarity. 

When the subsystem $A$ is a block of contiguous sites, the matrix $\Gamma_A$ is constructed as in Eq. \eqref{eq:Toeplitz}. However, we are interested in the case of disjoint intervals, where the construction is slightly more subtle. We consider in particular the case where $A=A_1\cup A_2$ consists of two non-complementary subsystems $A_1$ and $A_2$ of respective lengths $\ell_1$ and $\ell_2$ separated by a distance $d$, as shown in Fig. \ref{fig:Int}. Moreover, the length of the total subsystem $A$ is $\ell \equiv \ell_1+\ell_2$. We also mention that, since we work with periodic boundary conditions, one could in principle define two distances $d$. However, in the following we systematically consider the case where the subsystem $A$ is embedded in an infinite chain $L \to \infty$, and $d$ is the smallest distance between $A_1$ and $A_2$.

In this case, the correlation matrix $\Gamma_{A_1 \cup A_2}$ has the following block structure,
\beq
\label{eq:Gammad}
\Gamma_{A_1 \cup A_2}=\begin{pmatrix}\Gamma_{11}& \Gamma_{12}\\
\Gamma_{21}&\Gamma_{22}
\end{pmatrix}.
\eeq
Here, $\Gamma_{ab}$, $a,b=1,2$, are $2\ell_a \times 2\ell_b$ matrices that account for the correlations between subsystems $A_a$ and $A_b$,
\beq
\label{eq:blocks}
\begin{split}
\left[\Gamma_{11} \right]_{j,k}&=\Pi_{k-j},\quad  j,k=1,\cdots,\ell_1,\\
\left[\Gamma_{12} \right]_{j,k}&=\Pi_{(k+d+\ell_1)-j},\quad  j=1,\cdots,\ell_1, \quad k=1, \cdots, \ell_2,\\
\left[\Gamma_{21} \right]_{j,k}&=\Pi_{k-(j+d+\ell_1)},\quad  j=1,\cdots,\ell_2, \quad k=1, \cdots, \ell_1,\\
\left[\Gamma_{22} \right]_{j,k}&=\Pi_{k-j},\quad  j,k=1,\cdots,\ell_2,
\end{split}
\eeq
and the $2 \times 2$ blocks $\Pi_j$ are given in Eq. \eqref{eq:Toeplitz}. We note in particular that by definition, $\Gamma_{11}\equiv \Gamma_{A_1}$, and similarly for $A_2$.

Using Eqs. \eqref{eq:Trrhocorr}, \eqref{eq:Sn} and \eqref{eq:RMI}, we express the R\'enyi entropies and mutual information for disjoint intervals as
\begin{subequations}
\label{eq:SMIcorr}
\begin{equation}
\label{eq:Scorr}
    S_n^{A_1\cup A_2}=\frac{1}{2(1-n)}\Tr \log\left[\left(\frac{\id+\ir \Gamma_{A_1\cup A_2}}{2}\right)^n+\left(\frac{\id-\ir \Gamma_{A_1\cup A_2}}{2}\right)^n \right]
\end{equation}
and
\beq
\begin{split}
    I_n^{A_1:A_2}&=\frac{1}{2(1-n)}\Tr \log \left[\left(\frac{\id+\ir \Gamma_{11}}{2}\right)^n+\left(\frac{\id-\ir \Gamma_{11}}{2}\right)^n \right] \\ &+\frac{1}{2(1-n)}\Tr \log \left[\left(\frac{\id+\ir \Gamma_{22}}{2}\right)^n +\left(\frac{\id-\ir \Gamma_{22}}{2}\right)^n \right]\\ &- \frac{1}{2(1-n)}\Tr \log\left[\left(\frac{\id+\ir \Gamma_{A_1\cup A_2}}{2}\right)^n+\left(\frac{\id-\ir \Gamma_{A_1\cup A_2}}{2}\right)^n \right].
\end{split}
\eeq
\end{subequations}

We mention that the reduced density matrix $\rho_{A_1 \cup A_2}$ corresponding to disjoint subsystems is different in the spin-chain picture and in the free-fermion one \cite{atc-10,ip-10,fagotti2010entanglement}. This difference arises because of the fermionic string of the Majorana operators and boils down to the fact that the Jordan-Wigner transformation is non-local in terms of the spins. However, in the scaling limit, the entanglement dynamics is identical in both descriptions \cite{ac2-19} and hence we do not discuss this issue further.

\section{Quasiparticle conjecture for free fermions}\label{sec:QPP}

The quasiparticle picture is a semiclassical argument that allows one to derive quantitative predictions for the dynamics of entanglement in generic integrable models \cite{CC04,cc-05, ac-17,ac-18,c-20}. In this picture, a quantum system is prepared in an initial state that possesses an extensive amount of energy above the ground state of the Hamiltonian determining the time evolution, and hence the initial state acts as source of quasiparticles. The quasiparticles are emitted in pairs with opposite momenta $\pm k$ and move ballistically with a velocity $v_k$. The fundamental assumption is that quasiparticles emitted from the same point are entangled, and hence spread entanglement and correlations as they propagate. In contrast, quasiparticles emitted from different points are uncorrelated. Accordingly, the amount of entanglement between two subsystems at a given time is proportional to the number of entangled pairs they share.

\subsection{Single interval}
For a contiguous subsystem $A$ of length $\ell$ embedded in an infinite system in the scaling limit $\ell,t \to \infty$, with fixed ratio $t/\ell$, the entanglement entropy $S_1^A(t)$ is \cite{cc-05}
\beq
\label{eq:QPPEE}
S_1^A(t)=2 t \int_{2 v_kt<\ell} \dd k v_k s(k)+ \ell \int_{2 v_kt>\ell}\dd ks(k).
\eeq
Here, $s(k)$ is a function that depends on the momentum $k$ of each quasiparticle in the pair, and contains all the information on the initial state \cite{c-20}.

In free fermionic systems, the explicit form of $s(k)$ is
\beq
\label{eq:skfree}
s(k)=\frac{1}{2 \pi}(-n_k\log n_k-(1-n_k)\log(1-n_k)),
\eeq
where $n_k$ is the occupation probability of the mode $k$ in the stationary state. For the R\'enyi entropy $S_n(t)$, an expression analogous to \eqref{eq:QPPEE} holds, with 
\beq
\label{eq:snkfree}
s_n(k)=\frac{1}{2 \pi} \frac{1}{1-n}\log(n_k^n+(1-n_k^n))
\eeq
instead of $s(k)$. Introducing the function
\begin{equation}
\label{eq:hnx}
   h_n(x) \equiv \frac{1}{1-n}\log\left[\Big(\frac{1+x}{2}\Big)^n+\Big(\frac{1-x}{2}\Big)^n \right], 
\end{equation}
the R\'enyi entropies can be written as
\beq
\label{eq:QPPRE}
S_n^A(t)=\int \frac{\dd k}{2 \pi}h_{n}(2n_k-1)\min(\ell, 2 v_kt).
\eeq

In the case of interacting integrables models, Eq. \eqref{eq:QPPEE} needs to include a sum over the different species of quasiparticles \cite{ac-17}. Moreover, in that case it is a hard problem to derive the function $s(k)$, and the quasiparticle picture breaks down for R\'enyi entropies with $n\neq 1$, i.e. the simple prescription of Eq. \eqref{eq:snkfree} does not hold.

In the XY chain, Eq. \eqref{eq:QPPRE} becomes \cite{fc-08}
\beq
\label{eq:QPPRE_XY}
S_n^A(t)=\int_{-\pi}^{\pi} \frac{\dd k}{2 \pi}h_{n}(\cos \Delta_k)\min(\ell, 2 v_k t),
\eeq
where $v_k=|\epsilon_k'|$, and $\cos \Delta_k, \epsilon_k$ are defined in Eq. \eqref{eq:quenchparameters}. The quasiparticle conjecture of Eq. \eqref{eq:QPPRE_XY} is proved in Ref. \cite{fc-08}.

\subsection{Two disjoint intervals}

The quasiparticle picture also allows one to predict the time evolution of the entanglement between two disjoint blocks $A_1$ and $A_2$ \cite{ac-19,ac2-19} of lengths $\ell_1,\ell_2$ and separated by a distance $d$, as illustrated in Fig. \ref{fig:Int}. As for the case of one block, the entanglement between the two disjoint blocks is proportional to the number of shared pairs of quasiparticles over time. The counting exercise that leads to Eq. \eqref{eq:QPPEE} is modified by the presence of the two intervals, but its generalization is standard. The subsystem $A=A_1 \cup A_2$ is embedded in an infinite chain, and we work in the scaling limit $\ell_1,\ell_2,d,t \to \infty$ with fixed ratios $\ell_1/\ell, \ell_2/\ell, d/\ell$ and $t/\ell$, where $\ell=\ell_1+\ell_2$. Moreover, we introduce the function
\begin{equation}
\label{eq:Xi}
   \Xi_{\ell_1,\ell_2,d,t}(k) =\max(d, 2v_kt)+ \max(d+\ell, 2v_kt)- \max(d+\ell_1, 2v_k t)- \max(d+\ell_2, 2v_k t). 
\end{equation} 

For free fermionic systems, the R\'enyi mutual information reads \cite{ac2-19}
\begin{equation}
    \label{eq:QPPMI}
    I_n^{A_1:A_2}(t)=\int \frac{\dd k}{2 \pi} h_{n}(2n_k-1)\Xi_{\ell_1,\ell_2,d,t}(k),
\end{equation}
where $h_n(x)$ is defined in \eqref{eq:hnx}. We mention that the quench dynamics of logarithmic negativity can also be described by the quasiparticle picture\cite{ac2-19}, and the result is similar to Eq. \eqref{eq:QPPMI}. 

In the XY chain, the quasiparticle conjecture of Eq. \eqref{eq:QPPMI} reads 
\begin{equation}
    \label{eq:QPPMI_XY}
    I_n^{A_1:A_2}(t)=\int_{-\pi}^{\pi} \frac{\dd k}{2 \pi} h_{n}(\cos \Delta_k)\Xi_{\ell_1,\ell_2,d,t}(k).
\end{equation}

\subsection{Three intervals} 
\begin{figure}[t]
\centering
\includegraphics[width=0.8\textwidth]{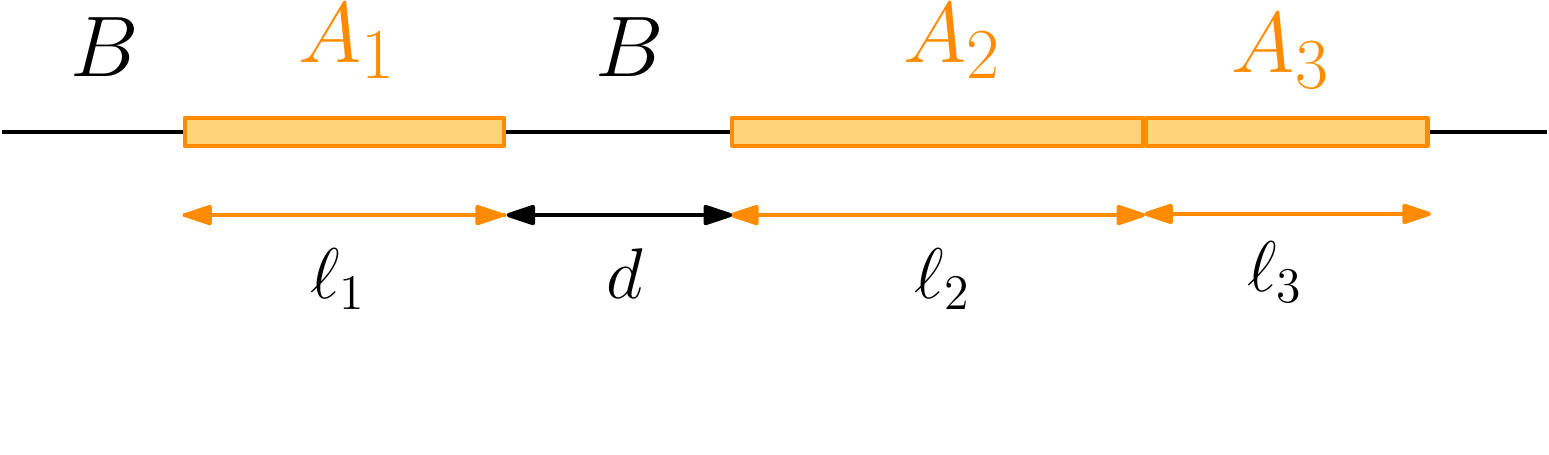}
\caption{The tripartite subsystem $A$ consists into three subsystems, $A_1$ and $A_2$ which are separated by a distance $d$, and $A_3$ that is adjacent to $A_2$.}
\label{fig:Tripartite}
\end{figure}

We consider the case where $A=A_1 \cup A_2 \cup A_3$ consists of three intervals of lengths $\ell_1,\ell_2,\ell_3$ with $\ell_1+\ell_2+\ell_3=\ell$. For simplicity, we assume that $A_1$ and $A_2$ are separated by a distance $d$, but $A_2$ and $A_3$ are adjacent. This geometry is illustrated in Fig. \ref{fig:Tripartite}. Combining Eqs. \eqref{eq:I3} and \eqref{eq:QPPMI}, we find
\begin{equation}
\begin{split}
\label{eq:I3QPP}
    I_n^{A_1:A_2:A_3}(t) &= \int \frac{\dd k}{2 \pi} h_{n}(2n_k-1)(\Xi_{\ell_1,\ell_2,d,t}(k)+\Xi_{\ell_1,\ell_3,d+\ell_2,t}(k)-\Xi_{\ell_1,\ell_2+\ell_3,d,t}(k)) \\
    &= 0
    \end{split}
\end{equation}
in the scaling limit where the lengths and time are infinite, but all the ratios are fixed. This simple calculation shows that, for free systems, the the tripartite information for the geometry of Fig.~\ref{fig:Tripartite} vanishes for all values of $t/\ell$ in the scaling limit. In particular, Eq.~\eqref{eq:I3QPP} holds for the XY chain. A direct generalization of Eq. \eqref{eq:I3QPP} with $s(k)$ instead of $h_{n}(2n_k-1)$, and a sum over the different species of quasiparticles, implies that the tripartite information  $I_1^{A_1:A_2:A_3}(t)$ vanishes identically for all interacting integrable systems. 

\subsection{Conjecture for the R\'enyi entropies of disjoint blocks in the XY chain}\label{sec:QPPconj}

In the case of two disjoint intervals, combining Eqs.~\eqref{eq:QPPRE_XY}, \eqref{eq:QPPMI_XY} and~\eqref{eq:RMI}, and using the fact that $A_1$ and $A_2$ are both contiguous blocks, the quasiparticle conjecture for the R\'enyi entropies reads \cite{fagotti2010entanglement}
\begin{equation}
\label{eq:QPP_conj_Sn_disj_XY}
    S_n^{A_1\cup A_2}(t)=\int_{-\pi}^{\pi} \frac{\dd k}{2 \pi}h_{n}(\cos \Delta_k)[\min(\ell_1, 2 v_k t)+\min(\ell_2, 2 v_k t)-\Xi_{\ell_1,\ell_2,d,t}(k)].
\end{equation}
It is direct to check that Eq. \eqref{eq:QPP_conj_Sn_disj_XY} reduces to Eq. \eqref{eq:QPPRE_XY} for $d=0$, as expected. In the following, we prove Eq. \eqref{eq:QPP_conj_Sn_disj_XY}, and hence Eqs. \eqref{eq:QPPMI_XY} and \eqref{eq:I3QPP} too, as byproducts. 

\subsection{Numerical checks}
\label{sec:numerics}

\begin{figure}
    \centering
    \includegraphics[scale=0.278]{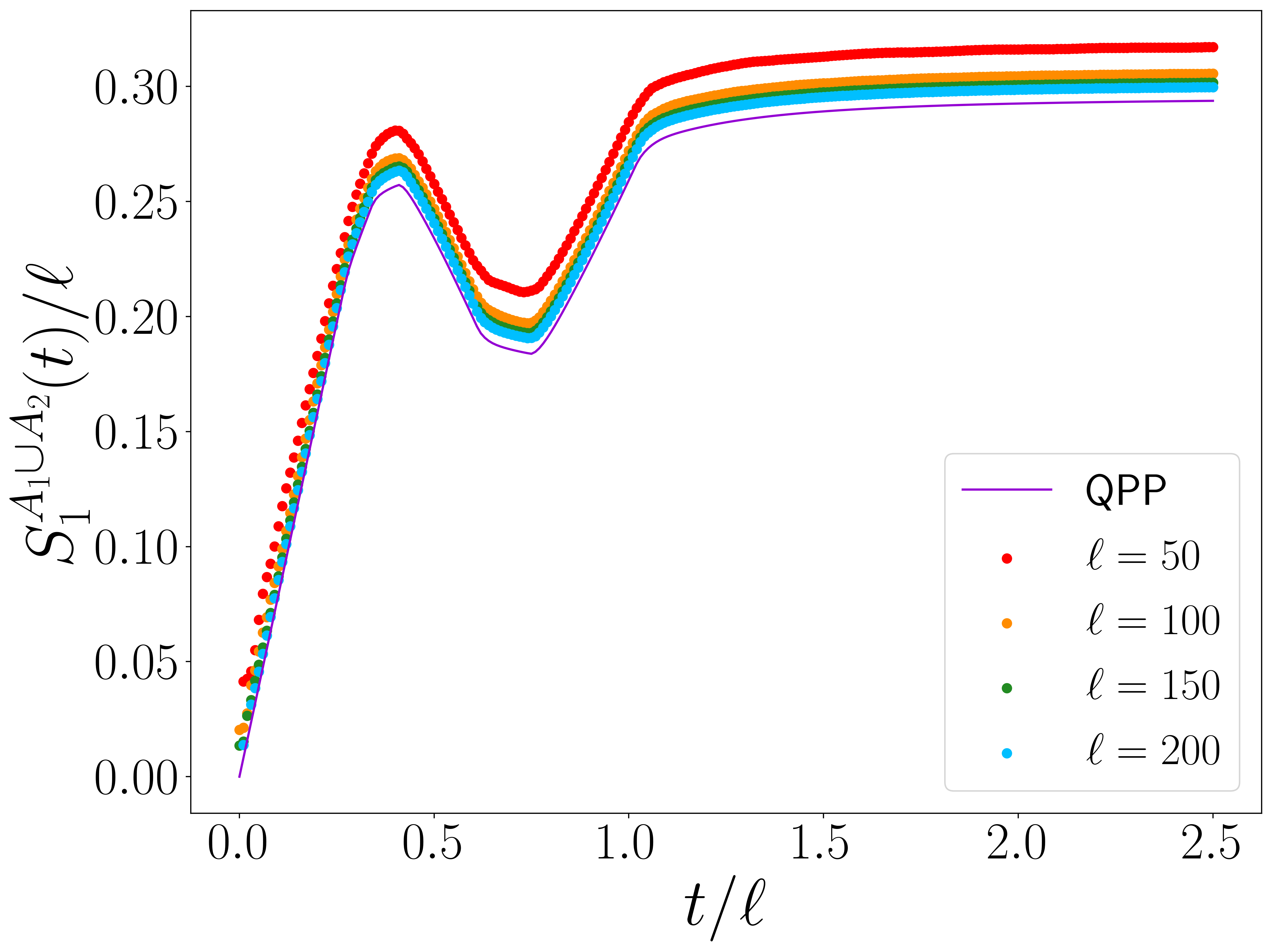}
     \includegraphics[scale=0.278]{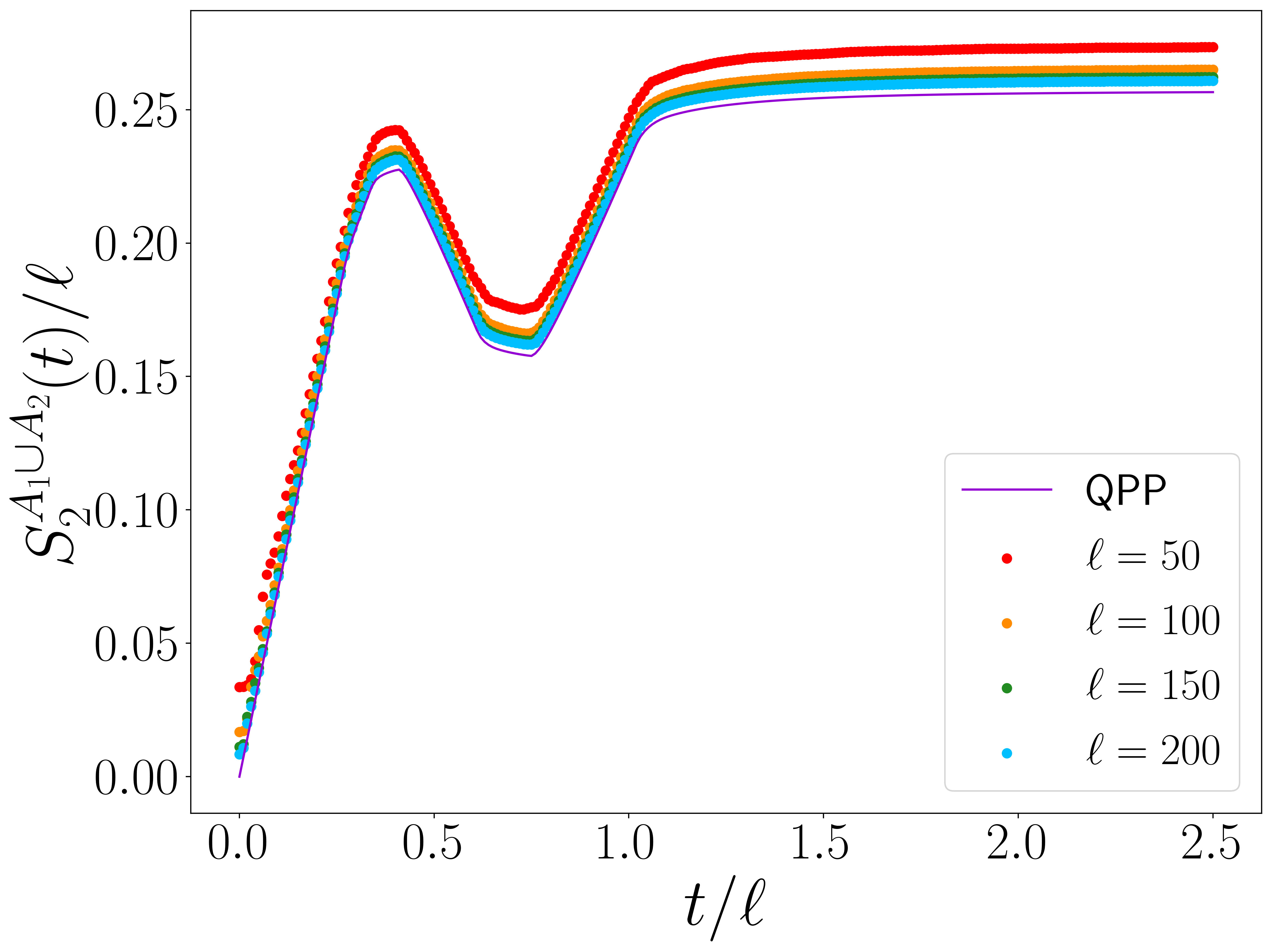}\\
     \includegraphics[scale=0.278]{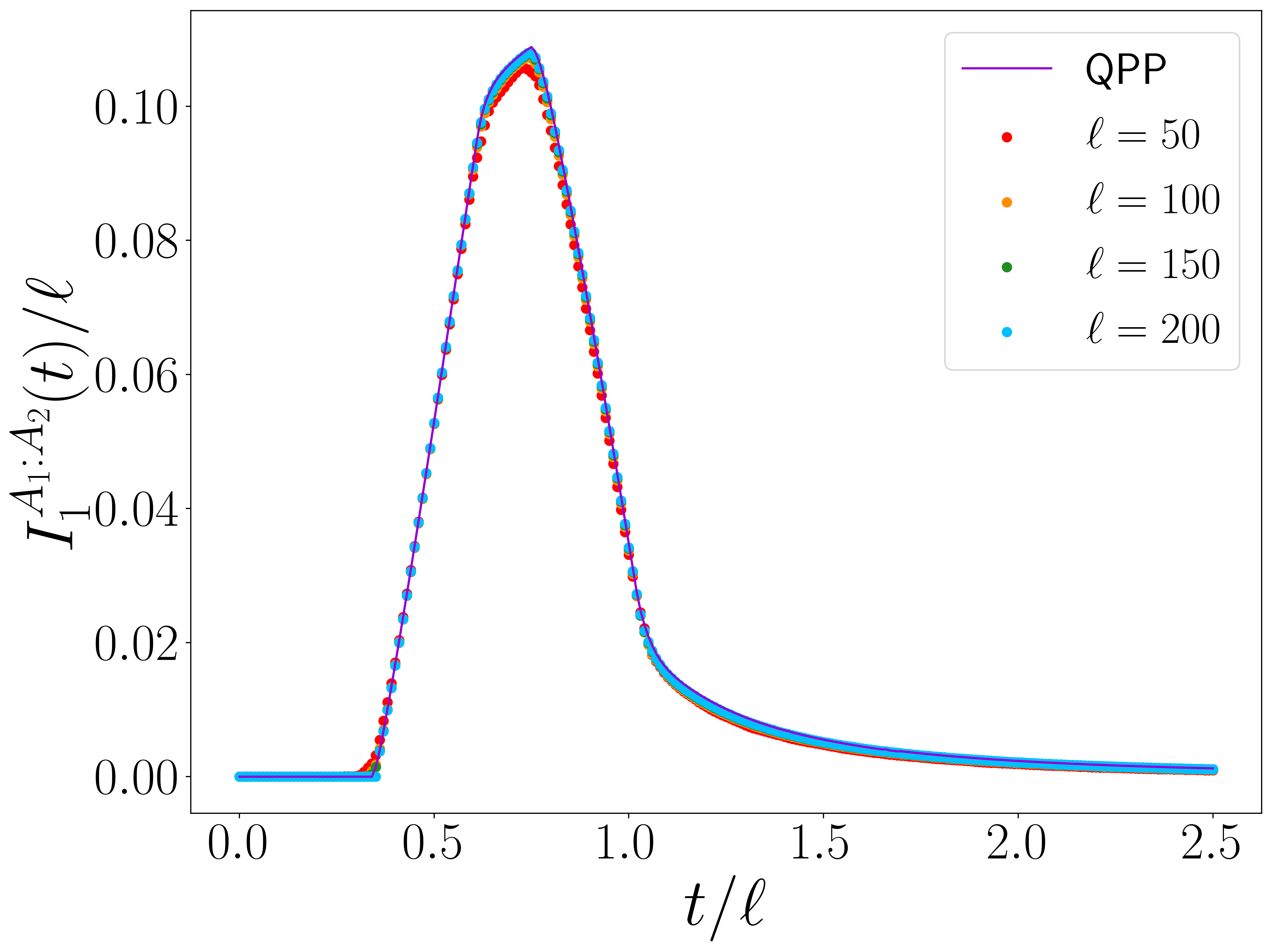}
     \includegraphics[scale=0.278]{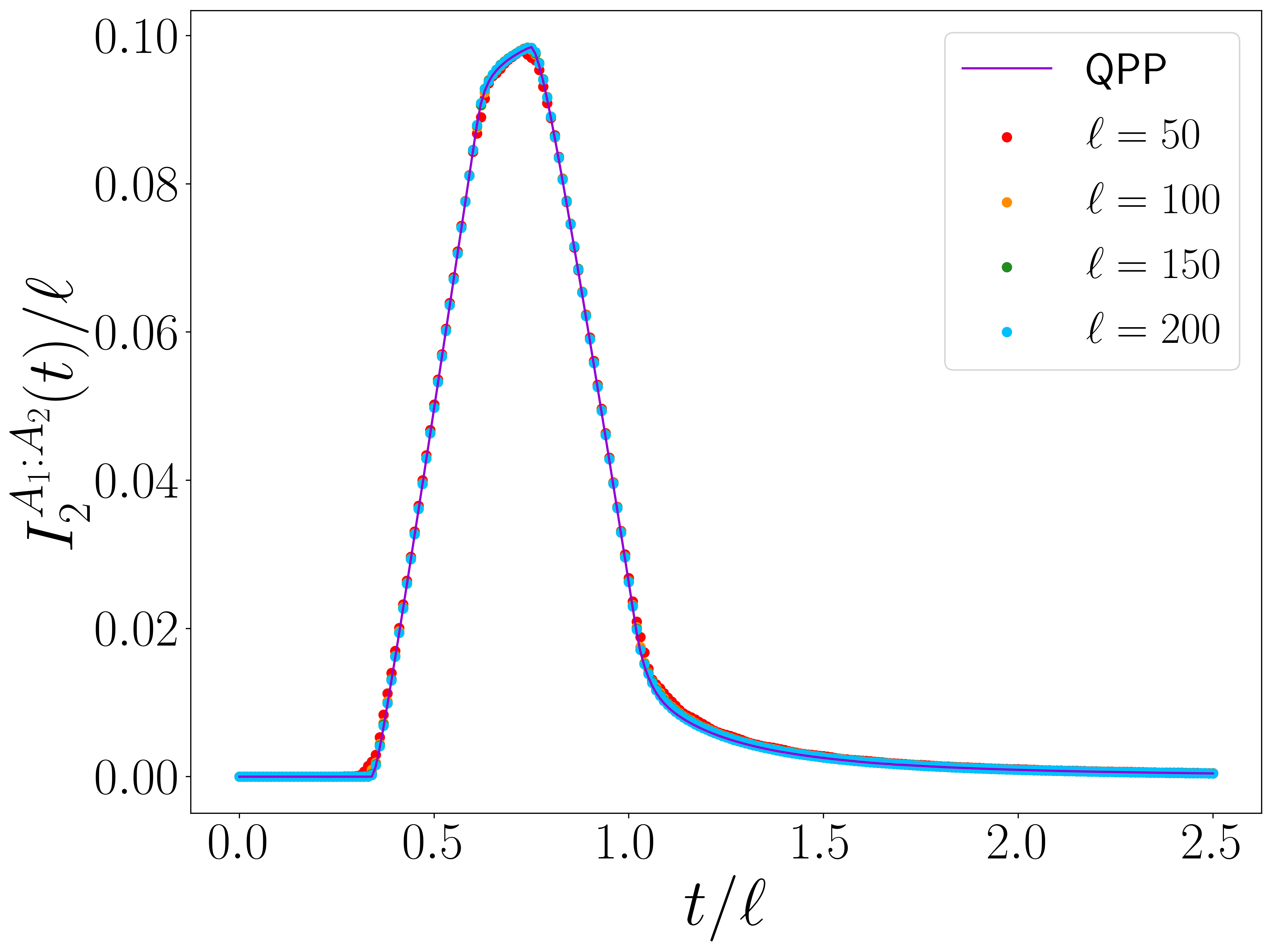}
    \caption{Entanglement entropies $S_1^{A_1 \cup A_2}(t)$ and $S_2^{A_1 \cup A_2}(t)$ (top) and mutual information $I_1^{A_1:A_2}(t)$ and $I_2^{A_1:A_2}(t)$ (bottom) as a function of $t/\ell$ for various values of $\ell$ with $\ell_1/\ell=0.4$ and $d/\ell=0.5$. The quench parameters are $h_0=0.1$, $\gamma_0=0.2$, $h=0.8$ and $\gamma=0.9$. The solid lines are the quasiparticle predictions of Eq. \eqref{eq:QPP_conj_Sn_disj_XY} (top) and Eq. \eqref{eq:QPPMI_XY} (bottom), and the symbols are obtained by exact diagonalization of the correlation matrices and Eq. \eqref{eq:SMIcorr}.}
    \label{fig:S_I}
\end{figure}

In this section, we compare the quasiparticle results of Eqs. \eqref{eq:QPP_conj_Sn_disj_XY} and \eqref{eq:QPPMI_XY} with numerical results that rely on the exact diagonalization of the correlation matrices from Eqs. \eqref{eq:Toeplitz} and \eqref{eq:Gammad} inserted in Eq. \eqref{eq:SMIcorr}. We show this comparison in Fig. \ref{fig:S_I} for $S_1^{A_1 \cup A_2}(t)$, $S_2^{A_1 \cup A_2}(t)$, $I_1^{A_1:A_2}(t)$ and $I_2^{A_1:A_2}(t)$ as a function of $t/\ell$ for fixed $d/\ell,\ell_1/\ell$ and quench parameters $\gamma_0,h_0,\gamma,h$. The solid lines are the quasiparticle results, whereas the symbols are the numerical ones for different values of $\ell$. The agreement is better for the mutual information, but the results for the entropies is convincing too, and it is in particular clear that the quasiparticle result is exact in the scaling limit, as expected. 

\section{Proof of the quasiparticle conjecture for two disjoint blocks}
\label{sec:proof}

In this section, we adapt the methods developed in Ref. \cite{fc-08} to prove Eq. \eqref{eq:QPP_conj_Sn_disj_XY}. The initial step is to recast Eq. \eqref{eq:Scorr} as a series. We introduce the Taylor expansion of the function $h_n(x)$,
\begin{equation}
\label{eq:hn_Taylor}
    h_n(x) = \sum_{j=0}^\infty c_n(2j)x^{2j},
\end{equation}
and find 
\begin{equation}
\label{eq:S_sum}
    S_n^{A_1\cup A_2}(t) = \frac 12 \sum_{j=0}^\infty c_n(2j) \Tr (\ir \Gamma_{A_1 \cup A_2}(t))^{2j}.
\end{equation}
We mention that the series in Eq. \eqref{eq:hn_Taylor} contains only even powers, because $h_n(x)=h_n(-x)$, and $h_n(x)$ is defined in Eq. \eqref{eq:hnx}. 

In the following, we thus focus on the computation of $\Tr (\ir \Gamma_{A_1 \cup A_2}(t))^{2j}$. For further convenience, we recast this trace as 
\begin{equation}
\label{eq:TrG_sep}
    \Tr (\ir \Gamma_{A_1 \cup A_2}(t))^{2j} \equiv \Tr (\ir \Gamma_{11}(t))^{2j}+\Tr (\ir \Gamma_{22}(t))^{2j}+T_{2j}(t),
\end{equation}
where we used the block structure described in Eq. \eqref{eq:Gammad}, and $T_{2j}(t)$ is a non-trivial combination of traces of products of $2j$ matrices from the set $\{\Gamma_{11}(t), \Gamma_{12}(t),\Gamma_{21}(t),\Gamma_{22}(t)\}$.

In the geometry we consider, both $A_1$ and $A_2$ are contiguous blocks of respective lengths $\ell_1$ and $\ell_2=\ell-\ell_1$, and their respective associated correlation matrices are $\Gamma_{11}$ and $\Gamma_{22}$. Hence the results of Ref. \cite{fc-08} directly apply to the first two terms of the right-hand side of Eq. \eqref{eq:TrG_sep}. In the scaling limit $\ell_1,\ell_2,t \to \infty$ with fixed ratios $\ell_1/\ell, \ell_2/\ell$ and $t/\ell$, we thus have 
\begin{equation}
    \label{eq:TrG11G22}
    \Tr (\ir \Gamma_{11}(t))^{2j}+\Tr (\ir \Gamma_{22}(t))^{2j} = 2\ell-2\int \frac{\dd k}{2 \pi}(1-(\cos \Delta_k)^{2j})[\min(\ell_1, 2 v_k t)+\min(\ell_2, 2 v_k t)]
\end{equation}
where we recall $v_k = |\epsilon_k'|$. 

It remains to investigate the term $T_{2j}(t)$ from Eq. \eqref{eq:TrG_sep} in the scaling limit $\ell_1,\ell_2,d,t \to \infty$ with fixed ratios $\ell_1/\ell, \ell_2/\ell, d/\ell$ and $t/\ell$. As expected, this term contains non-trivial dependence on the three lengths $\ell_1,\ell_2$ and $d$.

\subsection[Calculation of $T_2(t)$]{Calculation of $\boldsymbol{T_2(t)}$} For simplicity, we begin with the calculation of $T_2(t)$. We have
\begin{equation}
    \label{eq:T2}
    \begin{split}
    T_2&=-2 \Tr(\Gamma_{12}\Gamma_{21}) \\ &=2 \sum_{m_1=1}^{\ell_1}\sum_{m_2=1}^{\ell_2}\Big(2 f_{(m_2+\ell_1+d)-m_1}^2+g^2_{(m_2+\ell_1+d)-m_1}+g^2_{m_1-(m_2+\ell_1+d)}\Big)
    \end{split}
\end{equation}
where we used Eqs. \eqref{eq:Toeplitz} and \eqref{eq:blocks}, as well as $f_{-j}=-f_j$. In the notations, we dropped the explicit time dependence for clarity.

\paragraph{Summation formulas.}

First, we focus on terms of the form $f_j^2$ in Eq. \eqref{eq:T2}. With Eq. \eqref{eq:fg}, we find
\begin{equation}
    \label{eq:f2}
    \begin{split}
         f^2_{(m_2+\ell_1+d)-m_1}
         =\int_{-\pi}^{\pi}\int_{-\pi}^{\pi}\frac{\dd k_1 \dd k_2}{4 \pi^2} \eE^{-\ir(k_1-k_2)(m_2-m_1+\ell_1+d)}\sin\Delta_{k_1}\sin \Delta_{k_2}\sin(2 \epsilon_{k_1}t)\sin(2 \epsilon_{k_2}t).
    \end{split}
\end{equation}
With the identity
\begin{equation}
\label{eq:sumexp}
    \sum_{m=1}^{\ell}\eE^{\ir k m}=\eE^{\ir k \left(\frac{\ell+1}{2} \right)}\frac{\sin\left(\frac{\ell}{2}k \right)}{\sin\left( \frac{k}{2}\right)},
\end{equation}
we obtain
\begin{equation}
\label{eq:sumff}
\begin{split}
    \sum_{m_1=1}^{\ell_1}\sum_{m_2=1}^{\ell_2}f^2_{(m_2+\ell_1+d)-m_1}&=\int_{-\pi}^{\pi}\int_{-\pi}^{\pi} \frac{\dd k_1 \dd k_2}{4 \pi^2}\sin \Delta_{k_1}\sin \Delta_{k_2}\sin(2\epsilon_{k_1}t)\sin(2\epsilon_{k_2}t)\\
    &\times \left[\eE^{-\ir (k_1-k_2)\Big(\frac{\ell_1}{2}+\frac{\ell_2}{2}+d\Big)} \frac{\sin\left(\frac{\ell_1}{2}(k_1-k_2) \right) \sin\left(\frac{\ell_2}{2}(k_1-k_2) \right)}{\sin^2\left(\frac{(k_1-k_2)}{2} \right)}\right].
\end{split}
\end{equation}

Second, we consider terms of the form $g_j^2$ in Eq. \eqref{eq:T2}. We introduce
\begin{equation}
    \eta(k)=\cos \Delta_k+\ir \sin \Delta_k \cos (2 \epsilon_k t)
\end{equation}
and find
\begin{equation}
    \label{eq:g2}
        g^2_{(m_2+\ell_1+d)-m_1}=\int_{-\pi}^{\pi}\int_{-\pi}^{\pi}\frac{\dd k_1 \dd k_2}{4 \pi^2}\eE^{-\ir (k_1-k_2)(m_2-m_1+\ell_1+d)}\eE^{-\ir(\theta_{k_1}-\theta_{k_2})}\eta(k_1)\eta^*(k_2),
\end{equation}
where we used $\eta(k)=\eta^*(-k)$. Using Eq. \eqref{eq:sumexp}, we obtain
\begin{subequations}
\label{eq:sumgg}
\begin{equation}
\begin{split}
     \sum_{m_1=1}^{\ell_1}\sum_{m_2=1}^{\ell_2}g^2_{(m_2+\ell_1+d)-m_1}&=\int_{-\pi}^{\pi}\int_{-\pi}^{\pi}\frac{\dd k_1 \dd k_2}{4 \pi^2}\eE^{-\ir (\theta_{k_1}-\theta_{k_2})}\eta(k_1)\eta^*(k_2)\\
     &\times \left[\eE^{-\ir (k_1-k_2)\Big(\frac{\ell_1}{2}+\frac{\ell_2}{2}+d\Big)} \frac{\sin\left(\frac{\ell_1}{2}(k_1-k_2) \right) \sin\left(\frac{\ell_2}{2}(k_1-k_2) \right)}{\sin^2\left(\frac{k_1-k_2}{2} \right)}\right].
\end{split}
\end{equation}
Similarly, we find
\begin{equation}
\begin{split}
     \sum_{m_1=1}^{\ell_1}\sum_{m_2=1}^{\ell_2}g^2_{m_1-(m_2+\ell_1+d)}&=\int_{-\pi}^{\pi}\int_{-\pi}^{\pi}\frac{\dd k_1 \dd k_2}{4 \pi^2}\eE^{\ir (\theta_{k_1}-\theta_{k_2})}\eta^*(k_1)\eta(k_2)\\
     &\times \left[\eE^{-\ir (k_1-k_2)\Big(\frac{\ell_1}{2}+\frac{\ell_2}{2}+d\Big)} \frac{\sin\left(\frac{\ell_1}{2}(k_1-k_2) \right) \sin\left(\frac{\ell_2}{2}(k_1-k_2) \right)}{\sin^2\left(\frac{k_1-k_2}{2} \right)}\right].
\end{split}
\end{equation}
\end{subequations}

To proceed, we use the following identity in Eqs. \eqref{eq:sumff} and \eqref{eq:sumgg},
\begin{equation}
    \label{eq:sinratio}
    \frac{\sin\left(\frac{\ell}{2}(k_1-k_2) \right)}{\sin \left( \frac{k_1-k_2}{2}\right)}=\frac{\ell}{2}\frac{\left(\frac{k_1-k_2}{2}\right)}{\sin \left( \frac{k_1-k_2}{2}\right)}\int_{-1}^{1}\dd \zeta \cos \left(\frac{\ell \zeta}{2}(k_1-k_2) \right).
\end{equation}
Moreover, we replace $\eE^{\pm \ir (\theta_{k_1}-\theta_{k_2})}$ and $2 \sin\left(\frac{k_1-k_{2}}{2}\right)/(k_1-k_2)$ by one in the integrals, because they do not depend on $\ell,t$, and we anticipate that they are evaluated at $k_1=k_2$ in the stationary phase approximation. Finally, the purely imaginary terms arising from the products $\eta^*(k_1)\eta(k_2)$ and $\eta(k_1)\eta^*(k_2)$ in the integrals vanish by symmetry. We thus find
\begin{multline}
    \label{eq:T2final}
        T_2=\\ \ell_1 \ell_2 \int_{-\pi}^{\pi}\int_{-\pi}^{\pi}\frac{\dd k_1 \dd k_2}{4 \pi^2}\int_{-1}^{1}\int_{-1}^{1}\dd \zeta_1 \dd \zeta_2 \eE^{-\ir (k_1-k_2)\left(\frac{\ell}{2}+d \right)} \cos \left( \frac{\ell_1 \zeta_1}{2}(k_1-k_2)\right)\cos \left( \frac{\ell_2 \zeta_2}{2}(k_1-k_2)\right)\B_2(k_1,k_2),
\end{multline}
where 
\begin{equation}
    \label{eq:B2tilde}
    \B_2(k_1,k_2)=\cos \Delta_{k_1}\cos \Delta_{k_2}+ \sin \Delta_{k_1}\sin \Delta_{k_2}\cos(2 \epsilon_{k_1}t-2 \epsilon_{k_2}t).
\end{equation}

\paragraph{Integrals over $\boldsymbol{\zeta}$.} Let us focus on the integrals over $\zeta_1,\zeta_2$,
\begin{equation}
    Z_2 \equiv \int_{-1}^{1}\int_{-1}^{1}\dd \zeta_1 \dd \zeta_2\cos \left( \frac{\ell_1 \zeta_1}{2}(k_1-k_2)\right)\cos \left( \frac{\ell_2 \zeta_2}{2}(k_1-k_2)\right).
\end{equation}
Because the integrals are symmetric about $\zeta_{1,2}=0$ and the kernel is invariant under $\zeta_{1,2}\to -\zeta_{1,2}$, we have
\begin{equation}
   Z_2=  \int_{-1}^{1}\int_{-1}^{1}\dd \zeta_1 \dd \zeta_2\cos \left(\frac{\ell}{2}(x_2 \zeta_2-x_1 \zeta_1)(k_1-k_2)\right)
\end{equation}
with $x_j=\ell_j/\ell$, $j=1,2$. We introduce the change of variables
\begin{equation}
    \label{eq:rhovariable}
    \begin{split}
        \rho_0&=x_1\zeta_1\\
        \rho_1&=x_{2}\zeta_{2}-x_1\zeta_1, 
    \end{split}
\end{equation}
and obtain
\begin{equation}
\label{eq:Z2_rho}
    Z_2= \frac{\ell^2}{\ell_1 \ell_2} \int_{-x_1}^{x_1}\dd \rho_0 \int_{-x_2-\rho_0}^{x_2-\rho_0} \dd \rho_1 \cos \left(\frac{\ell}{2}\rho_1(k_1-k_2) \right). 
\end{equation}

The bounds of integration depend on $\rho_0$, whereas the kernel do not, and it is possible to reduce this double integral as a single integral over $\rho_1$. For a generic function $J(\rho_1)$, we have
\begin{subequations}
\label{eq:Frho1}
\begin{equation}
    \int_{-x_1}^{x_1}\dd \rho_0 \int_{-x_2-\rho_0}^{x_2-\rho_0}\dd \rho_1 J(\rho_1)=\int_{\mathbb{R}}\dd \rho_1 J(\rho_1)\Omega_{12}(\rho_1),
\end{equation}
where
\begin{equation}
    \Omega_{12}(\rho_1)=\max[0, \min(x_1,x_2+\rho_1)+\min(x_1,x_2-\rho_1)].
\end{equation}
\end{subequations}
We stress that Eq. \eqref{eq:Frho1} is a non-trivial generalization of a similar result used in Ref. \cite{fc-08} for the case of a single block.

Combining Eqs. \eqref{eq:T2final}, \eqref{eq:Z2_rho} and \eqref{eq:Frho1}, we find 
\begin{equation}
\label{eq:T2_rho}
    T_2= \frac{\ell^2}{2\pi} \int_{-\pi}^\pi \frac{\dd k}{2 \pi}\left[ \int_{[-\pi, \pi]\times \mathbb{R}}\dd k_1 \dd \rho_1 \eE^{-\ir (k_1-k)\left(\frac{\ell}{2}+d \right)} \B_2(k_1,k) \eE^{ \frac{\ir \ell\rho_1}{2}(k_1-k)}  \Omega_{12}(\rho_1)\right]
\end{equation}
where we used the symmetry about $0$ of the integral over $\rho_1$ to replace $\cos \left(\frac{\ell}{2}\rho_1(k_1-k_2) \right)$ by an exponential.  

\paragraph{Multidimensional stationary phase approximation.} We study the asymptotic behaviour of the two-fold integral over $k_1$ and $\rho_1$ in Eq. \eqref{eq:T2_rho} using multidimensional stationary phase approximation. This method can be used to evaluate integrals of the form
\begin{equation}
    \label{eq:SPAintegral}
    I(\ell)= \int_{\mathbb{R}^n}\dd x \ G(x)\eE^{\ir \ell F(x)},
\end{equation}
in the limit $\ell \to \infty$. In the case where there is only one point $x_0 \in \mathbb{R}^n$ such that $\nabla F(x_0)=0$, we have
\begin{equation}
    \label{eq:SPA}
    \lim_{\ell \rightarrow \infty}I(\ell)= G(x_0)\eE^{\ir \ell F(x_0)}\left( \frac{2 \pi}{\ell}\right)^{\frac{n}{2}}|\det \mbox{Hess}(F(x_0))|^{-\frac{1}{2}}+o(\ell^{-\frac{n}{2}}).
\end{equation}

Using the definition \eqref{eq:B2tilde} of $\B_2$, the integral in Eq. \eqref{eq:T2_rho} has two distinct terms, one is time-independent, and the second contains all the time dependence. We study these two terms separately. 

For the time-independent term, the two-fold integral can be written in the form of Eq. \eqref{eq:SPAintegral} with 
\begin{equation}
\begin{split}
    F(k_1, \rho_1) &= -(k_1-k)\left(\frac{1}{2}+ \frac{d}{\ell}\right)+ \frac{\rho_1}{2}(k_1-k), \\
    G(k_1,\rho_1) &= \cos \Delta_{k_1}\cos \Delta_{k} \Omega_{12}(\rho_1).
    \end{split}
\end{equation}
The stationary point is $x_0 = (k, 1+\frac{2d}{\ell})$. It is direct to verify that $\Omega_{12}(1+\frac{2d}{\ell})=0$, and hence the time-independent term does not contribute in the scaling limit. 

For the time dependent term, we write $\cos(2 \epsilon_{k_1}t-2 \epsilon_{k_2}t)$ from $\B_2$ in exponential form, and hence we have two terms with opposite phases. Both integrals have the form of Eq. \eqref{eq:SPAintegral}
with 
\begin{equation}
\begin{split}
    F_t(k_1, \rho_1) &= -(k_1-k)\left(\frac{1}{2}+ \frac{d}{\ell}\right)+ \frac{\rho_1}{2}(k_1-k) \pm \frac{2t}{\ell}(\epsilon_{k_1}-\epsilon_k) , \\
    G_t(k_1,\rho_1) &= \sin \Delta_{k_1}\sin \Delta_{k} \Omega_{12}(\rho_1).
    \end{split}
\end{equation}
The stationary points are $x_0 = (k, 1+\frac{2d}{\ell} \mp \frac{4t}{\ell}\epsilon_{k}')$. We have $|\det \mbox{Hess}(F_t(x_0))|=2^{-2}$, and hence Eqs.~\eqref{eq:SPA} and \eqref{eq:T2_rho} yield 
\begin{equation}
    T_2(t) = \ell \int_{-\pi}^\pi \frac{\dd k}{2 \pi} (1-(\cos \Delta_k)^2)\left[\Omega_{12}\Big( 1+\frac{2d}{\ell} + \frac{4t}{\ell}\epsilon_{k}'\Big)+\Omega_{12}\Big( 1+\frac{2d}{\ell} - \frac{4t}{\ell}\epsilon_{k}'\Big)\right]. 
\end{equation}
Similarly as for the time-independent term, $\Omega_{12}\Big( 1+\frac{2d}{\ell}+a\Big)=0$ for $a\geqslant 0$, and hence  we recast $T_2(t)$ as
\begin{equation}
     T_2(t) = \ell \int_{-\pi}^\pi \frac{\dd k}{2 \pi} (1-(\cos \Delta_k)^2)\Omega_{12}\Big( 1+\frac{2d}{\ell} - \frac{4t}{\ell}v_k\Big)
\end{equation}
with $v_k=|\epsilon_k'|$. A direct calculation gives $ \Omega_{12}\Big( 1+\frac{2d}{\ell} - \frac{4t}{\ell}v_k\Big)=2/\ell  \ \Xi_{\ell_1,\ell_2,d,t}(k)$, where $\Xi_{\ell_1,\ell_2,d,t}(k)$ is defined in \eqref{eq:Xi}, and hence we conclude
\begin{equation}
\label{eq:T2_final}
     T_2(t) = 2 \int_{-\pi}^\pi \frac{\dd k}{2 \pi} (1-(\cos \Delta_k)^2) \ \Xi_{\ell_1,\ell_2,d,t}(k).
\end{equation}

\subsection[Calculation of $T_{2j}(t)$]{Calculation of $\boldsymbol{T_{2j}(t)}$}

In this section, we generalize the calculations of the previous one to the case of $T_{2j}(t)$, where $j$ is a positive integer, and we recall that $T_{2j}(t)$ is defined in Eq.~\eqref{eq:TrG11G22}. It is a sum of traces of various combinations of products of $\Gamma_{ab}$ matrices. For convenience, we first focus on one certain type of sets of matrices $\Gamma_{ab}$, namely 
\begin{equation}
\label{eq:Gamma_rlj}
    \Gamma_{r,l,j} = \{\underbrace{\Gamma_{12},\Gamma_{21}}_{r-1 \ \text{times}}, \Gamma_{12}, \underbrace{\Gamma_{22},\dots,\Gamma_{22}}_{2j-2r-l \ \text{times}}, \Gamma_{21},\underbrace{\Gamma_{11}, \dots,\Gamma_{11}}_{l \ \text{times}}  \}
\end{equation}
where $r,l$ are positive integers that satisfy $r > 0$ and $2j-2r-l\geqslant 0$. The matrix at position $i=1,\dots,2j$ in the set $\Gamma_{r,l,j}$ is $\Gamma_{a_i b_i}$. We also define $X_{r,l,j}$ as the set where the entry at position $i=1,\dots,2j$ is the ratio $X_{r,l,j}(i)=\ell_{a_i}/\ell$. Finally, introduce the quantity $C_{r,l,j}$ as the trace of the product of all the matrices in $\ir \Gamma_{r,l,j}$,
\begin{equation}
    \label{eq:Crljdef}
     C_{r,l, j} \equiv (-1)^{j}\Tr[(\Gamma_{12}\Gamma_{21})^{r-1}\Gamma_{12} \Gamma_{22}^{2j-2r-l}\Gamma_{21}\Gamma_{11}^l],
\end{equation}
where we introduced the sign $(-1)^j=\ir^{2j}$, because in the definition of $T_{2j}$, all the $\Gamma_{ab}$ matrices are multiplied by $\ir$. As an example, for $r=2,l=1,j=3$, we have 
\begin{equation}
\begin{split}
    \Gamma_{2,1,3} &= \{\Gamma_{12},\Gamma_{21},\Gamma_{12},\Gamma_{22},\Gamma_{21},\Gamma_{11}\},\\[.3cm]
    X_{2,1,3} &= \Big\{\frac{\ell_1}{\ell},\frac{\ell_2}{\ell},\frac{\ell_1}{\ell},\frac{\ell_2}{\ell},\frac{\ell_2}{\ell},\frac{\ell_1}{\ell}\Big\}, \\[.3cm]
    C_{2,1,3} &= -\Tr[\Gamma_{12}\Gamma_{21}\Gamma_{12} \Gamma_{22}\Gamma_{21}\Gamma_{11}].
    \end{split}
\end{equation}

\paragraph{Computation of $\boldsymbol{C_{r,l,j}}$.}
 Using similar calculation as in the previous section, we find 
\begin{equation}
\label{eq:Crlj}
    C_{r,l, j}=\frac{\ell_1^{r+l}\ell_2^{2j-r-l}}{2^{2j-1}}\int \left(\prod_{i=1}^{2j} \frac{\dd k_i}{2 \pi}\right)\left(\prod_{i=1}^{2j} \dd \zeta_i \right)\eE^{\ir \left( \frac{\ell}{2}+d\right)\left(\sum_{i=1}^{2r-1}(-1)^ik_i+k_{2j-l} \right)}\B_{2j} \prod_{i=1}^{2j}\cos \left( \frac{\ell}{2}x_i\zeta_i(k_i-k_{i-1})\right),
\end{equation}
where $x_i=X_{r,l,j}(i)$, see previous paragraph, and we assume a periodicity of $2j$ in the indices, namely $k_0=k_{2j}$. The function $\B_{2j}$ is a non-trivial generalization of the $B$ function from Ref. \cite{fc-08}. We consider the set of integers $\lambda^{(2j)}=\{1,2,\dots,2j\}$, and we denote by $\lambda_{2p}^{(2j)} \in \lambda^{(2j)}$ a subset which contains $2p\leqslant 2j$ of these integers, and $\bar{\lambda}_{2p}^{(2j)}$ is its complement, $\lambda_{2p}^{(2j)} \cup \bar{\lambda}_{2p}^{(2j)}=\lambda^{(2j)}$. We impose that both sets are ordered, namely their elements are strictly increasing. Moreover, for $i \in \lambda_{2p}^{(2j)}$, the function $s(i)$ gives the position of $i$ in the set. For example, for $j=2$, one choice of subset is $\lambda_{2}^{(4)}=\{2,3\}$, and we have $s(2)=1$ and $s(3)=2$. With these definitions, we find
\begin{equation}
\label{eq:B2j}
    \B_{2j} = \sum_{p=0}^{j} \sum_{\lambda_{2p}^{(2j)}} (-1)^{|\lambda_{2p}^{(2j)}|+p}\left[\cos \Bigg( 2t \sum_{u \in \lambda_{2p}^{(2j)}} (-1)^{s(u)} \epsilon_{k_u}  \Bigg)\prod_{u \in \lambda_{2p}^{(2j)}} \sin \Delta_{k_u} \prod_{v \in \bar{\lambda}_{2p}^{(2j)}} \cos \Delta_{k_v} \right]
\end{equation}
with $|\lambda_{2p}^{(2j)}|=\sum_{i \in \lambda_{2p}^{(2j)}} i$. As examples, $\B_2$ is given in Eq. \eqref{eq:B2tilde}, and $\B_4$ is 
\begin{equation}
\begin{split}
    \B_4 &= \cos \Delta_{k_1}\cos \Delta_{k_2}\cos \Delta_{k_3}\cos \Delta_{k_4} \\
    &+ \cos(2t(\epsilon_{k_1}-\epsilon_{k_2})) \sin \Delta_{k_1}\sin \Delta_{k_2} \cos \Delta_{k_3}\cos \Delta_{k_4} \\
    &- \cos(2t(\epsilon_{k_1}-\epsilon_{k_3})) \sin \Delta_{k_1}\sin \Delta_{k_3} \cos \Delta_{k_2}\cos \Delta_{k_4} \\
    &+ \cos(2t(\epsilon_{k_1}-\epsilon_{k_4})) \sin \Delta_{k_1}\sin \Delta_{k_4} \cos \Delta_{k_2}\cos \Delta_{k_3} \\
    &+ \cos(2t(\epsilon_{k_2}-\epsilon_{k_3})) \sin \Delta_{k_2}\sin \Delta_{k_3} \cos \Delta_{k_1}\cos \Delta_{k_4} \\
    &- \cos(2t(\epsilon_{k_2}-\epsilon_{k_4})) \sin \Delta_{k_2}\sin \Delta_{k_4} \cos \Delta_{k_1}\cos \Delta_{k_3} \\
    &+ \cos(2t(\epsilon_{k_3}-\epsilon_{k_4})) \sin \Delta_{k_3}\sin \Delta_{k_4} \cos \Delta_{k_1}\cos \Delta_{k_2} \\
    &+ \cos(2t(\epsilon_{k_1}-\epsilon_{k_2}+\epsilon_{k_3}-\epsilon_{k_4}))\sin \Delta_{k_1}\sin \Delta_{k_2} \sin \Delta_{k_3}\sin \Delta_{k_4}. 
    \end{split}
\end{equation}

Going back to Eq. \eqref{eq:Crlj}, using the symmetry about $0$ of the $\zeta_i$ integrals, we transform the product of cosines into the cosine of a sum as follows
\begin{equation}
    \label{eq:Aproduct}
    \prod_{i=1}^{2j}\cos \left( \frac{\ell}{2}x_i\zeta_i(k_i-k_{i-1})\right)=\cos \left(\frac{\ell}{2}\sum_{i=1}^{2j-1}(x_{i+1}\zeta_{i+1}-x_i\zeta_i)(k_{i}-k_{2j}) \right).
\end{equation}
We stress that this equation is not an identity and only holds in the integrals. To proceed, we introduce the change of variables
\begin{equation}
    \label{eq:rhovariableN}
    \begin{split}
        \rho_0&=x_1\zeta_1\\
        \rho_i&=x_{i+1}\zeta_{i+1}-x_i\zeta_i, \qquad i=1, \dots,2j-1,
    \end{split}
\end{equation}
and obtain
\begin{equation}
\label{eq:Crljrho}
C_{r,l,j}=\ell\left(\frac{\ell}{2} \right)^{2j-1} \int \left(\prod_{i=1}^{2j}\frac{\dd k_i}{2 \pi} \right)\left(\prod_{i=0}^{2j-1}\dd \rho_i \right)\eE^{\ir \left( \frac{\ell}{2}+d\right)\left(\sum_{i=1}^{2r-1}(-1)^i k_i+k_{2j-l} \right)}\B_{2j}\cos \left( \frac{\ell}{2}\sum_{i=1}^{2j-1}\rho_i (k_i-k_{2j})\right).
\end{equation}
Here, the integral over $\rho_i$ is between $\left( -x_{i+1}-\sum_{m=0}^{i-1}\rho_m \right)$ and $ \left( x_{i+1}-\sum_{m=0}^{i-1}\rho_m \right)$. Similarly as for the case $j=1$, the kernel does not depend on $\rho_0$, and we have 
\begin{multline}
   \label{eq:CrljOmega}
   C_{r,l,j}= \ell \left( \frac{\ell}{4 \pi}\right)^{2j-1} \int_{-\pi}^{\pi} \frac{\dd k}{2 \pi}\Bigg[\int_{[-\pi,\pi]^{2j-1}\times \mathbb{R}^{2j-1} }  \left(\prod_{i=1}^{2j-1}\dd k_i\dd \rho_i\right) \\ \times \eE^{\ir \left(\frac{\ell}{2}+d \right)\left(\sum_{i=1}^{2r-1}(-1)^i k_i+k_{2j-l} \right)} \B_{2j} 
   \cos \left( \frac{\ell}{2}\sum_{i=1}^{2j-1}\rho_i(k_i-k)\right)\Omega_{12}(\{\rho_i\}) \Bigg ],
\end{multline}
where we set $k_{2j}\equiv k$, and
\begin{equation}
    \label{eq:Omega}
    \Omega_{12}(\{\rho_i\})=\max\left[0, \min_{\substack{w=1,\cdots, 2j-1}} \left( x_1,x_{w+1}+\sum_{m=1}^w\rho_m\right)+ 
\min_{\substack{w=1,\cdots, 2j-1}}\left( x_1,x_{w+1}-\sum_{m=1}^w\rho_m\right) \right].
\end{equation}

We treat the $(4j-2)$-fold integral in Eq.~\eqref{eq:CrljOmega} using the multidimensional stationary phase approximation discussed in the previous section. Crucially, the only term in $\B_{2j}$ that contributes in the scaling limit is the one where the time-dependent factor contains energies whose signs and indices match those of the $k_i$ in the overall the phase $\eE^{\ir \left(\frac{\ell}{2}+d \right)\left(\sum_{i=1}^{2r-1}(-1)^i k_i+k_{2j-l} \right)}$ in Eq. \eqref{eq:CrljOmega}. This time-dependent term is $(-1)^l\cos\left(2t \sum_{i=1}^{2r-1}(-1)^i \epsilon_{k_i}+2t \epsilon_{k_{2j-l}}\right)(\cos \Delta_k)^{2j-2r}(\sin \Delta_k)^{2r}$, where we anticipated that all the momenta are evaluated at $k_i=k$ in the stationary phase approximation. The sign $(-1)^l$ is a direct consequence of the definition of $\B_{2j}$, see Eq. \eqref{eq:B2j}. The stationary phase calculation is similar to the case $j=1$, and we find 
\begin{equation}
\label{eq:CrljSPA}
    C_{r,l,j}=(-1)^{l}\int_{-\pi}^{\pi} \frac{\dd k}{2 \pi} (\cos \Delta_k)^{2j-2r}(\sin \Delta_k)^{2r} \Xi_{\ell_1,\ell_2,d,t}(k).
\end{equation}

\paragraph{Reconstruction of $\boldsymbol{T_{2j}(t)}$.} Thus far, we have investigated the scaling behaviour of a very specific type of trace of product of $\Gamma_{ab}$ matrices, namely the quantity $C_{r,l,j}$ defined in Eq. \eqref{eq:Crljdef}. Of course, in $T_{2j}$, there are many more contributions, arising from different ways of arranging the $\Gamma_{ab}$ matrices. However, in the scaling limit, the traces of products of matrices taken from a same set but arranged in different orders are equal. We stress that this is not true in finite size, because the blocks of the matrices do not commute. Hence, in the scaling limit, we can write the general formula
\begin{equation}
\label{eq:T2nC}
T_{2j} =  \sum \limits_{r=1}^j \sum \limits_{l=0}^{2j-2r} \alpha_{r,l,j} C_{r,l,j},
\end{equation}
where $\alpha_{r,l,j}$ is the number of ways of arranging the matrices from the set $\Gamma_{r,l,j}$ defined in Eq. \eqref{eq:Gamma_rlj} such that the resulting product is a square matrix of size $\ell_1 \times \ell_1$ or $\ell_2 \times \ell_2$. A simple combinatorial argument yields 
\begin{equation}
\label{eq:alpha}
\alpha_{r,l,j} = \frac{2r j}{(2j-r-l)(r+l)} \begin{pmatrix}
2j-r-l \\
r
\end{pmatrix}
\begin{pmatrix}
r+l \\
r
\end{pmatrix}. 
\end{equation}

Crucially, the coefficients $\alpha_{r,l,j}$ satisfy \cite{boyack, pbc-21bis}
\begin{equation}
\label{eq:sumalpha}
    \sum_{l =0}^{2j - 2r}(-1)^{l}\alpha_{r,l,j}=2 \binom{j}{r}.
\end{equation}

Finally, using the binomial relation
\begin{equation}
    \sum_{r=1}^j \binom{j}{r}(\cos \Delta_k)^{2j-2r}(\sin \Delta_k)^{2r}=1-(\cos \Delta_k)^{2j}
\end{equation}
together with Eqs. \eqref{eq:CrljSPA}, \eqref{eq:T2nC} and \eqref{eq:sumalpha}, we get
\begin{equation}
    \label{eq:T2jQPP}
    T_{2j}(t)=2  \int \frac{\dd k}{2 \pi}(1-(\cos \Delta_k)^{2j}) \Xi_{\ell_1,\ell_2,d,t}(k)
\end{equation}
in the scaling limit. This result appears to be a straightforward generalization of Eq. \eqref{eq:T2_final}, but the proof is not. 

\subsection{Final resummation and conclusion of the proof}

To conclude the proof, we need to perform the resummation of Eq.~\eqref{eq:S_sum}. First, we note that since $h_n(1)=0$, the sum of all the Taylor coefficients vanishes, $\sum_{j=0}^\infty c_n(2j)=0$. Hence, the terms that do not depend on the summation index $j$ in the quantities we sum over do not contribute. Combining Eqs. \eqref{eq:hn_Taylor}, \eqref{eq:S_sum}, \eqref{eq:TrG_sep}, \eqref{eq:TrG11G22} and \eqref{eq:T2jQPP}, we find 
\begin{equation}
    S_n^{A_1\cup A_2}(t)=\int_{-\pi}^{\pi} \frac{\dd k}{2 \pi}h_{n}(\cos \Delta_k)[\min(\ell_1, 2 v_k t)+\min(\ell_2, 2 v_k t)-\Xi_{\ell_1,\ell_2,d,t}(k)],
\end{equation}
as claimed in Sec. \ref{sec:QPPconj}, Eq.~\eqref{eq:QPP_conj_Sn_disj_XY}. 

As advertised, a direct corollary is that Eq.~\eqref{eq:QPPMI_XY} holds for the mutual information. Indeed, following the same strategy as for the entropies of disjoint blocks, we have 
\begin{equation}
\begin{split}
      I_n^{A_1: A_2}(t) &= - \frac 12 \sum_{j=0}^\infty c_n(2j) T_{2j}(t) \\
      &=\int_{-\pi}^{\pi} \frac{\dd k}{2 \pi}h_{n}(\cos \Delta_k)\Xi_{\ell_1,\ell_2,d,t}(k).
      \end{split}
\end{equation}

Finally, this result also implies that the tripartite information identically vanishes for all times, see Eq. \eqref{eq:I3QPP}, in agreement with the results of Ref. \cite{maric2022universality}.

\

\section{Conclusion}
\label{sec:conclusion}

In this manuscript, we provide the exact \textit{ab initio} derivation of the formula for the dynamics of the entanglement entropies for two disjoint blocks in the spin-1/2 XY chain in a transverse magnetic field, after a global quantum quench. This result, given by Eq. \eqref{eq:QPP_conj_Sn_disj_XY}, also provides a proof that the quasiparticle picture for the mutual information holds in that model, see Eq. \eqref{eq:QPPMI_XY}. Our approach generalizes the one used in Eq. \cite{fc-08} for the case of a single interval. In particular, it consists in expressing the R\'enyi entropies of two disjoint blocks $S_n^{A_1\cup A_2}(t)$ as a sum of terms containing the traces of even powers of the correlation matrix $\Gamma_{A_1\cup A_2}(t)$. The evaluation of such terms
amounts to that of sum of traces of different combinations of products of blocks of the correlation matrix, whose final expression is obtained via multidimensional stationary phase approximation in the scaling limit. Another direct implication of our calculation is that the tripartite information identically vanishes at all times for the XY chain, and more generally for systems where the quasiparticle picture holds.

There are some natural generalizations of our results. First, it would be interesting to see if the exact methods developed in Refs. \cite{ac2-22,caceffo2022entanglement} can be adapted to address the quench dynamics of the negativity in the XY chain. This would prove the relation between the logarithmic negativity and the R\'enyi mutual information with index $n=1/2$ after a quench \cite{ac2-19}. Second, our calculations can probably be adapted to the case of three intervals. This would for example allow us to generalize Eq. \eqref{eq:I3QPP} to study the tripartite information for disjoint intervals, and see if it also vanishes identically in that case. Finally, our results can certainly be adapted to derive exact expressions for the recently introduced entanglement asymmetry \cite{ares2022entanglement} and R\'enyi fidelities \cite{parez2022symmetry} in the context of disjoint intervals.

\section*{Acknowledgements}

GP holds a CRM-ISM postdoctoral fellowship and acknowledges support from the Mathematical Physics Laboratory of the CRM. He also thanks SISSA for hospitality during the early stages of this project. RB~acknowledges support from the Croatian Science Foundation (HrZZ) project No. IP-2019-4-3321. We thank Pasquale Calabrese and Vincenzo Alba for useful discussions and comments on the manuscript. We also thank Rufus Boyack for sharing his proof of Eq. \eqref{eq:sumalpha}.

\providecommand{\href}[2]{#2}\begingroup\raggedright\endgroup

\end{document}